\documentclass[journal,twoside,web]{ieeecolor}
\usepackage{tmi}
\usepackage{cite}
\usepackage{amsmath,amssymb,amsfonts}
\usepackage{algorithmic}
\usepackage{graphicx}
\usepackage{textcomp}
\usepackage{siunitx}
\usepackage{url}
\usepackage{makecell}
\usepackage{caption}
\usepackage{epstopdf}

\usepackage{glossaries}

\newcommand{\NTX}{N_{\mathrm{TX}}}
\newcommand{\NEL}{N_{\mathrm{CH}}}
\newcommand{\NAX}{N_{\mathrm{FT}}}
\newcommand{\NS}{N_{\mathrm{SC}}}
\newcommand{\NX}{N_{\mathrm{X}}}
\newcommand{\NZ}{N_{\mathrm{Z}}}
\newcommand{\wvfm}[2]{\phi_{#1}\left(#2 \right)}
\newcommand{\allwvfm}{\Phi}
\newcommand{\allscatpos}{P_s}
\newcommand{\scatpos}{\mathbf{p}}
\newcommand{\allscatamps}{\mathbf{a}}
\newcommand{\scatamp}{a}

\newcommand{\tzdel}{ \Psi }
\newcommand{\ttx}{\tau_{\mathrm{TX}}}
\newcommand{\trx}{\tau_{\mathrm{RX}}}
\newcommand{\tz}{\tau_0}
\newcommand{\elw}{\mathrm{elw}}
\newcommand{\opt}{\xi}

\newcommand{\systempar}{S}
\newcommand{\probegeometry}{P_{\text{probe}}}
\newcommand{\elpos}[1]{{\mathbf{p}_{#1}}}
\newcommand{\attenuation}{\mu}
\newcommand{\optvar}[1]{\xi_{#1}}
\newcommand{\elgain}{\gamma}
\newcommand{\plotradius}{r}

\newcommand{\allaperture}{\mathbf{u}}
\newcommand{\aperture}{u}
\newcommand{\beamformedsample}{z}
\newcommand{\directivity}{b}
\newcommand{\attspreadtx}{\alpha_{TX}}
\newcommand{\attspreadrx}{\alpha_{RX}}
\newcommand{\attabsorption}{a_{abs}}
\newcommand{\ax}{\mathrm{ft}}
\newcommand{\el}{\mathrm{ch}}
\newcommand{\tx}{\mathrm{tx}}
\newcommand{\pixel}{{pix}}
\newcommand{\eltx}{\mathrm{txel}}
\newcommand{\elrx}{\mathrm{rxel}}
\newcommand{\scat}{s}
\newcommand{\txapod}{\Upsilon}
\newcommand{\initialtime}{\tau_{\text{init}}}

\newacronym{das}{DAS}{delay-and-sum}
\newacronym{rf}{RF}{radio frequency}
\newacronym{method}{INFER}{INverse grid-Free Estimation of Reflectivities}
\newacronym{gcnr}{gCNR}{generalized contrast-to-noise ratio}
\newacronym{tgc}{TGC}{time gain compensation}
\newacronym{adc}{ADC}{analog-to-digital conversion}
\newacronym{tof}{TOF}{time-of-flight}
\newacronym{mv}{MV}{minimum-variance}
\newacronym{fwhm}{FWHM}{full width half maximum}
\newacronym{dmas}{DMAS}{delay-multiply-and-sum}
\newacronym{iq}{I/Q}{in-phase quadrature}
\newacronym{sgd}{SGD}{stochastic gradient descent}
\newacronym{admm}{ADMM}{alternating direction method of multipliers}
\newacronym{red}{RED}{regularization by denoising}
\newacronym{sa}{SA}{synthetic aperture}
\newacronym{mse}{MSE}{mean squared error}
\newacronym{nlm}{NLM}{nonlocal means}
\newacronym{ista}{ISTA}{iterative shrinkage-thresholding algorithm}

\DeclareMathOperator*{\argmin}{arg\,min}

\def\BibTeX{{\rm B\kern-.05em{\sc i\kern-.025em b}\kern-.08em
    T\kern-.1667em\lower.7ex\hbox{E}\kern-.125emX}}
\title{Off-Grid Ultrasound Imaging by Stochastic Optimization}
\author{Vincent van de Schaft, \IEEEmembership{Member, IEEE}, Oisín Nolan, \IEEEmembership{Member, IEEE}, Ruud van Sloun, \IEEEmembership{Member, IEEE}
\thanks{This work was performed within the IMPULSE framework of the Eindhoven MedTech Innovation Center (e/MTIC, incorporating Eindhoven University of Technology and Philips Research), including a PPS supplement from the Dutch Ministry of Economic Affairs and Climate Policy.}
\thanks{Vincent van de Schaft, Oísin Nolan, and Ruud J. G. van sloun are with the Department of Electrical Engineering, Eindhoven University of Technology, 5612 AZ Eindhoven, The Netherlands (email: v.v.d.schaft@tue.nl; o.nolan@tue.nl; r.j.g.v.sloun@tue.nl)}
\thanks{Special thanks to Dr. Oudom Somphone and Dr. Sofiane Decombas-Deschamps at Philips Research Paris for their advice and support during this project.}
\thanks{This work has been submitted to the IEEE for possible publication. Copyright may be transferred without notice, after which this version may no longer be accessible.}
}

\begin{document}

\maketitle

\begin{abstract}
Ultrasound images formed by delay-and-sum beamforming are plagued by artifacts that only clear up after compounding many transmissions. Some prior works pose imaging as an inverse problem. This approach can yield high image quality with few transmits, but requires a very fine image grid and is not robust to changes in measurement model parameters. We present \gls{method}, an off-grid and stochastic algorithm that solves the inverse scattering problem in ultrasound imaging. Our method jointly optimizes for the locations of the gridpoints, their reflectivities, and the measurement model parameters such as the speed of sound. This approach allows us to use significantly fewer gridpoints, while obtaining better contrast and resolution and being more robust to changes in the imaging target and the hardware. The use of stochastic optimization enables solving for multiple transmissions simultaneously without increasing the required memory or computational load per iteration. We show that our method works across different imaging targets and across different transmit schemes and compares favorably against other beamforming and inverse solvers. The source code and the dataset to reproduce the results in this paper are available at \url{www.github.com/vincentvdschaft/off-grid-ultrasound}.
\end{abstract}


\section{Introduction}
\noindent Ultrasound imaging allows physicians to observe internal tissues in real-time without the need for ionizing radiation, but it provides poor image quality compared to other imaging modalities such as computed tomography imaging, magnetic resonance imaging, or X-ray imaging.

This poor image quality is largely the result of artifacts that are inherent in the way we map ultrasound signals into an image. In conventional B-mode ultrasound imaging, an image is formed by applying some variant of \gls{das} beamforming. The \gls{das} algorithm selects the \gls{rf} data sample from each transducer channel that contains the peak of the backscattered signal from the target location. These samples are then summed to produce an estimate of the signal strength coming from that location. This signal strength is known as the reflectivity. A flaw in this approach is that samples do not only contain signal from the target location, but also from all other points sharing the same \gls{tof}. \gls{das} nevertheless works quite well because the signals from the target location sum coherently, while the signals from other locations tend to sum incoherently. This incoherent summing suppresses undesired signal components. As we sum over more channels and compound more transmissions, the error in the reflectivity estimate gets smaller. However, as any real-world ultrasound acquisition can only have a finite number of channels and transmissions, we end up with B-mode images that show clutter in hypoechoic regions, and strong, spatially correlated speckle patterns in areas with tissue, obscuring the underlying anatomy.

To reduce these artifacts, more advanced adaptive beamforming techniques have been proposed that actively suppress the unwanted components based on the signal content. The \gls{mv} beamformer, for instance, minimizes the signal power under the constraint that the power in the target direction remains equal to 1 \cite{synnevag_adaptive_2007}. Spatial coherence-based beamformers such as the \gls{dmas} beamformer\cite{matrone_delay_2015} and Short-Lag-Spatial-Coherence beamformer \cite{lediju_short-lag_2011} apply a weighting to the received signals based on the coherence across the aperture. These algorithms achieve impressive improvements in clutter rejection, but have inherent limitations because they process the pixels in the image independently. Pixel-wise solutions can never fully cancel out unwanted signals because unwanted signals can be coherent by chance.

To move beyond pixel-wise solutions, there has been a growing interest in generating ultrasound images from \gls{rf} data by posing imaging as an inverse problem. Inverse-problem-based methods do not generate the image from a weighted sum of samples. Instead, inverse-problem-based methods model a forward process that produces measurement data based on a map of material properties like reflectivity or on a set of scatterers. The challenge is then to invert this model to recover the input that would have resulted in the observed data according to the model. This approach makes it possible to find a joint solution over all pixels in the image resulting in a better reconstruction.
The promise of inverse-problem-based approaches is that they may be able to produce ultrasound images that are free of sidelobe clutter and have significantly better contrast and resolution, all while requiring very few transmissions. This would thus not only improve image quality, but also increase attainable frame rates.
Several authors have proposed different inverse problem formulations over the years, each posing the inverse problem slightly differently and using different methods for optimization.

Firstly, there is the choice in how to model the forward process. In most prior works the authors model the transmit waveform as an impulse, a triangular pulse, or a rectangular pulse \cite{besson_compressed_2016, besson_ultrafast_2018, szasz_beamforming_2016, ozkan_inverse_2018, congzhi_wang_plane-wave_2015}. Every sample in the \gls{rf} data is then a (weighted) sum over all pixels with similar \gls{tof}. This formulation is referred to as \textit{regularized beamforming}\cite{besson_ultrafast_2018}. Regularized beamforming has a very sparse forward process matrix, which makes the problem more computationally tractable. It also retains speckle statistics in the images, which is relevant for various downstream tasks. However, because regularized beamforming formulations disregard all information present in the known waveform shape, they are not able to fully exploit the information present in the data.

The alternative is to model the actual transmit waveform in the forward process model. This formulation is referred to as the \textit{inverse scattering} problem. Both Schiffner et al.\cite{schiffner_fast_2011} and David et al.\cite{david_time_2015} define linear forward processes for the inverse scattering problem. They compute a large matrix containing the \gls{rf} data response to every scatterer in an image grid.

A second design choice is how to improve the conditioning of the often ill-posed inverse problem. Many works use an $l_1$-norm regularizer as a sparsity prior \cite{schiffner_compressed_2012, chernyakova_fourier-domain_2014, david_time_2015, besson_compressed_2016, besson_ultrafast_2018, congzhi_wang_plane-wave_2015, nicolet_simultaneous_2022, zhang_diffusion_2023}. This approach is especially suitable for imaging wire targets and acquisitions of diluted microbubbles, as the target images are sparse. For many other applications, target images are not sparse. The problem then remains to find a proper sparsifying basis.
To that end, Goudarzi et al. showed that the \gls{red} framework can be applied to inverse-problem-based ultrasound imaging\cite{goudarzi_inverse_2022}. Ozkan et al. apply a combination of heuristic regularizers\cite{ozkan_inverse_2018}. Zhang et al. apply denoising diffusion restoration models to solve the regularized beamforming problem under an image-based prior\cite{zhang_diffusion_2023}.

What all of these methods have in common is that they define a grid of scatterer positions and then solve for the amplitudes of the scatterers on this grid that satisfy the forward model. The problem with this is that actual sources of backscatter do not necessarily conform to the predefined grid positions. Therefore we either need to define a very fine grid, leading to increased computational- and memory-requirements, or we need to accept the offset in position, leading to artifacts in the resulting image, as noted by David et al. \cite{david_time_2015}.\\
Additionally, the aforementioned methods all use a static forward process model. Some recent work proposed to optimize for model parameters during imaging. Simson et al. developed a differentiable \gls{das} beamformer that adapts the speed of sound to minimize phase error \cite{simson_differentiable_2023}. This approach results in better image quality while also providing a speed of sound, showing that accurate and flexible estimation of physical parameters can improve image quality. Conversely, inverse-problem-based imaging methods that use pre-computed matrices of scatterer responses cannot change anything about the forward process without re-computing the full matrix.

In this work we make the following contributions:
\begin{itemize}
    \item We present \gls{method}, an off-grid inverse-problem-based method for the inverse scattering problem in ultrasound imaging. Our method defines a matrix-free model and applies stochastic gradient descent over the \gls{rf} samples to jointly optimize for the backscatter intensities, the scatterer locations, and the other parameters that determine the forward process model like the speed of sound and attenuation.
    \item We show that our method works across different transducers, imaging targets, and acquisition schemes, both in a tissue mimicking phantom and in \textit{in-vivo} acquisitions.

\end{itemize}
The rest of the paper is outlined as follows: In section \ref{sec:methods} we provide a detailed overview of the design considerations and implementation of \gls{method}. 
In section \ref{sec:experimental_setup} we discuss how we evaluate our method and compare it to several baseline algorithms. We then present some results on a tissue mimicking phantom and on several \textit{in-vivo} acquisitions in section \ref{sec:results}. We close with a discussion and conclusion in section \ref{sec:discussion}, and \ref{sec:conclusion} respectively.

\section{Methods}
\begin{figure}
    \centering
    \includegraphics[width=\linewidth]{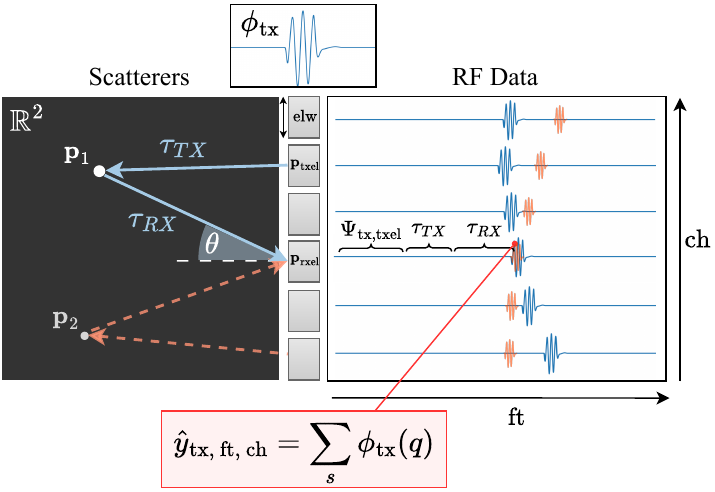}
    \caption{Visualization of the predicted channel data based on the scatterer positions and transmit waveform. The orange and blue responses correspond to two different scatterers.}
    \centering
    \label{fig:forward_model}
\end{figure}

\label{sec:methods}

\subsection{Overview}
\label{sec:overview}
\noindent Characterizing all components of the ultrasound acquisition process with sufficient accuracy to enable reliable inverse-solving \textit{in-vivo} is a daunting task. Not only due to the complexity, but also because many aspects of the process such as properties of imaging targets, and the exact characteristics of the hardware are not known in advance. Element sensitivities are specific to a transducer and the speed of sound is specific to each patient and anatomy. Therefore, to be able to properly solve the inverse problem we must solve not only for the target image, but also for the unknown parameters of the forward model.
\\
We achieve this by defining various model parameters as functions of free optimization variables. This way we can find a solution that satisfies a physical model where the model parameters are free to move within predefined ranges.
We further assume that the responses of scatterers sum linearly to produce the final \gls{rf} data. The response of a single scatterer, however, does not need to be a linear function of model parameters. The only requirement is that the function that maps the model parameters to the scatterer response is differentiable or can be approximated by a differentiable function. As such there is a lot of freedom in defining the model. Under these constraints, we can then apply stochastic gradient descent over \gls{rf} data samples to optimize for the scatterers and model parameters.

\subsection{Definitions and notation}
\label{sec:definitions}
\noindent Throughout the paper, we will sometimes use multiple-character names for variables in equations to improve readability. To prevent any confusion with products of single character variables, we will always denote multiplication with $a\cdot b$. We will introduce variables with their name followed by a unit in square brackets e.g. $c\in \mathbb{R}_{>0}$ [\si{\meter\per\second}]. Vectors are denoted with boldfaced symbols $\mathbf{v}$.

\subsubsection{Target anatomy}
We model the target anatomy as a collection of $\NS\in \mathbb{N}_{>0}$ scatterers, each with a position $\scatpos_s\in\mathbb{R}^2$, and a backscatter intensity $\scatamp\in \mathbb{R}_{\geq 0}$. The full tissue model is then described by a matrix containing all scatterer positions $\allscatpos\in\mathbb{R}^{2\times \NS}$, and a vector containing all scatterer amplitudes $\allscatamps\in \mathbb{R}^{\NS}$.
We assume an average speed of sound in the tissue $c$ [\si{\meter\per\second}]. We also assume an average attenuation coefficient in the tissue $\attenuation$ [\si{\decibel\per\centi\meter\per\mega\hertz}].

\subsubsection{Transducer}
The transducer is modelled as a set of $\NEL\in \mathbb{N}_{>0}$ transducer elements, each with a position $\elpos{\el}\in\mathbb{R}^2$ [\si{\meter}], where $\el\in\{0, \dots, \NEL-1\}$. We combine these positions into a matrix $\probegeometry\in \mathbb{R}^{2\times \NEL}$. We denote the width of the elements by $\elw$ [\si{\meter}]. We make the assumption that the elements are positioned along the $x$-axis with the emitting surface pointing in the positive $z$-direction, but the model can be easily adjusted to work with curved transducers by adding a direction per element.

\subsubsection{Transmit scheme}
The settings of the ultrasound system that determine the wave fields that are emitted will be referred to as the \textit{transmit scheme}. An acquisition consists of $\NTX\in \mathbb{N}_{>0}$ transmissions. The transmit scheme is defined by its three parts.
Firstly, there are the \textit{transmit delays}. The shape of the transmitted wavefront is determined by the time at which each of the elements fires relative to each other. We define $\tzdel\in \mathbb{R}_{\geq 0}^{\NTX\times\NEL}$ [\si{\second}] to be a matrix containing the transmit delays for every element for every transmit. In $\tzdel$, the element $\tzdel_{\tx, \el}$ is the time instant at which element $\el$ starts transmitting for transmission $\tx$. For each transmit we define $t=0$ such that $\min_\el \tzdel_{\tx, \el} = 0$ i.e. $t=0$ is when the first element starts firing. After $t=0$ the system waits for $\initialtime$ seconds before beginning to capture the $\NAX\in \mathbb{N}_{>0}$ samples with sampling frequency $f_s$.
Then, there is the transmit waveform. This is the pulse that every individual element emits starting at the corresponding transmit delay time. We will denote the transmit waveform of transmission $\tx\in\{0, \dots, \NTX-1\}$ as $\wvfm{\tx}{t}$. The original transmit waveform will be band-pass filtered because it passes through the transducer element and electronics twice. For convenience we define $\wvfm{\tx}{t}$ to be the already two-way-band-pass-filtered waveform. We let $\allwvfm$ denote the set of all used transmit waveforms.
Finally, the transmit apodization $\txapod\in \mathbb{R}_{\geq 0}^{\NTX\times\NEL}$[-] is a per-element gain in transmit that is set to $0$ for elements that do not fire in a transmit event.

\subsection{Forward measurement model}
\subsubsection{Core model}
\label{sec:forward_model}
\noindent Here we introduce the model we use for the measurement process. The measurement process maps a set of system parameters and the scatterer configuration to a data cube of predicted \gls{rf} data $Y\in\mathbb{R}^{\NTX\times\NAX\times \NEL}$, where $\NAX$ is the number of samples along the fast-time axis. Let $\systempar=\{\elgain, \elw, \attenuation, \initialtime, c, \opt_a, \opt_b\}$ be the set of all measurement model parameters as will be introduced later. Let $W$ be a matrix of white Gaussian noise samples. We can then express $Y$ as

\begin{equation}
    Y = f_Y(\allscatpos, \allscatamps, \systempar) + W.
    \label{eq:rf_full}
\end{equation}

\noindent A sample with fast time index $\ax$ of a receiving element $\el$ to all scatterers after they are excited by the wave field from a transmit event $\tx$ is then given by

\begin{equation}
    y_{\tx, \ax, \el} = f_y(\tx, \ax, \el, \allscatpos, \allscatamps, \systempar)+w.
    \label{eq:rf_single}
\end{equation}
We now need to define this function $f_y$.

We assume that a scatterer in the tissue causes a response in every channel of the transducer when excited by a transmitted wave. This response is just the transmit waveform shifted by the transmit delay and the travel delay.
The travel delay for the receiving transducer element $\elrx$ located at $\mathbf{p}_\elrx$ is just the time it takes the wavefront to get from transmitting element $\eltx$ to the scatterer and then back to element $\elrx$, which we will denote $\ttx$, and $\trx$ respectively. The received signal at element $\elrx$ thus becomes
\begin{equation}
    \phi_\tx(t-\ttx-\trx-\tzdel_{\tx,\eltx}).
    \label{eq:waveform}
\end{equation}
The travel time from element position $\elpos{\el}$ to scatterer position $\scatpos_\scat$ is simply the Euclidean distance divided by the speed of sound. The travel time to and from the scatterer can thus be computed as
\begin{align}
    \ttx & = \frac{\left\|\elpos{\eltx}-\scatpos_\scat\right\|}{c}, & \trx & = \frac{\left\|\elpos{\elrx}-\scatpos_\scat \right\|}{c}.
\end{align}
The sample with index $\ax$ is recorded at $t=\frac{\ax}{f_s}+\initialtime$, which means that we substitute $t=\frac{\ax}{f_s}+\initialtime$ in (\ref{eq:waveform}) to find the response of the scatterer in sample $\ax$.

We now have an expression for the response of a single transmitting element and a single scatterer. To compute the combined response over all transmitting elements and all scatterers, we simply sum the individual responses:
\begin{equation}
    y_{\tx, \ax, \el} = \sum_{\eltx}\sum_{\scat} \phi_\tx(\frac{\ax}{f_s}+\initialtime-\ttx-\trx-\tzdel_{\tx,\eltx})+w.
    \label{eq:model_allelements}
\end{equation}
\noindent Fig. \ref{fig:forward_model} shows a schematic visualization of the forward process.

\begin{figure}
    \centering
    \includegraphics{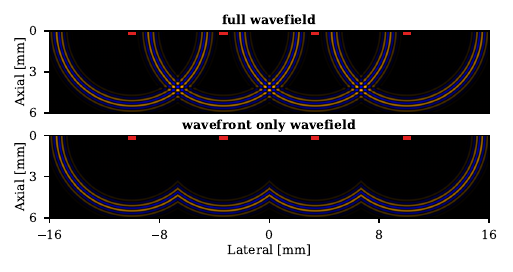}
    \caption{Illustration of the wave field simulated by the full model corresponding to (\ref{eq:model_allelements}) (top) and the wavefront only model corresponding to (\ref{eq:model_wavefrontonly}) (bottom).}
    \label{fig:wavefields}
\end{figure}

To reduce the computational load we can approximate the wave field by simulating not the contributions of all transmitting elements, but only the global wavefront. We do this by considering only the first wave that reaches the scatterer as:
\begin{align}
     & y_{\tx, \ax, \el} = \sum_{\scat}
    \phi_\tx(q)+w\nonumber                                                                         \\
     & q = \frac{\ax}{f_s}+\initialtime-\trx- \min_{\eltx}\left( \ttx + \tzdel_{\tx,\eltx}\right).
    \label{eq:model_wavefrontonly}
\end{align}
This simplification improves the runtime at the cost of a potentially larger model error. We will refer to the model in (\ref{eq:model_allelements}) as the \textit{full model} and to the model in (\ref{eq:model_wavefrontonly}) as the \textit{wavefront only} model. The difference between the simulated wave fields is illustrated in Fig. \ref{fig:wavefields}.

The definitions in (\ref{eq:model_allelements}) and (\ref{eq:model_wavefrontonly}) are the core of the model. In the remainder of this section we will introduce additional factors that modify the model to include various physical phenomena.

\subsubsection{Element directivity}
\label{sec:methods_directivity}
\noindent The sensitivity of transducer elements is dependent on the angle of incidence of the incoming wavefront. We will refer to this sensitivity as the element directivity. We assume an incoming planar wavefront of a single frequency, coming in with an angle $\theta$ [\si{\radian}] with respect to the normal pointing out of the element. Let $\lambda$ [\si{\meter}] be the wavelength of the incoming wave given its frequency and the speed of sound in the medium. Let $\elw$[\si{\meter}] be the width of the transducer element. The directivity can then be modeled as:
\begin{equation}
    \directivity(\theta) = \text{sinc} \left(\frac{ \elw\cdot \sin (\theta)}{\lambda}\right)\cdot \cos \left( \theta\right).
    \label{eq:directivity}
\end{equation}
This model was originally derived by Selfridge et al.\cite{selfridge_theory_1980}. This model comes with two simplifying assumptions: Firstly, because (\ref{eq:directivity}) assumes a planar wavefront it is only valid in the far field of the transducer element (note: not the far field of the array). Secondly, an actual pulse used in ultrasound imaging is a wideband signal. We will, however, only use the center frequency of the transducer to compute $\lambda$.

This directivity $\directivity(\theta)$ is applied both in transmit and receive. This means that when a scatterer has angle $\theta_{TX}$ with the transmitting element and angle $\theta_{RX}$ with the receiveing element, the \gls{rf} response will be scaled with a factor $\directivity(\theta_{TX})\cdot\directivity(\theta_{RX})$.

\subsubsection{Element gain}
\label{sec:methods_gain}
\noindent Not all elements receive the \gls{rf} signal equally well. Transducer elements tend to deteriorate or even break completely as a transducer gets older. In other cases the transducer may not be coupled with the skin correctly or otherwise blocked. These effects lead to differences in sensitivity. For these cases we model a scalar gain per element $\elgain\in \mathbb{R}^{\NEL}_{\geq 0}$ [\si{-}]. We include this gain per element as a scaling factor in the model of the \gls{rf} response.

\subsubsection{Attenuation}
\label{sec:methods_attenuation}
\noindent We model two types of attenuation of the wave: attenuation due to absorption $\attabsorption\in \mathbb{R}_{\geq 0}$[-], and attenuation due to the spread of the wave as it propagates to the scatterer and back, denoted as $\attspreadtx\in \mathbb{R}_{\geq 0}$[-], and $\attspreadrx\in \mathbb{R}_{\geq 0}$[-] respectively. The corresponding travel distances are $d_{TX}$ [\si{\meter}], and $d_{RX}$ [\si{\meter}]. Let $f_c$ [\si{\hertz}] be the carrier frequency. The attenuation due to absorption is then defined as
\begin{equation}
    \attabsorption = 10^{-\frac{\attenuation}{20}\cdot f_c\cdot 10^{-6}\cdot 100\cdot (d_{TX}+d_{RX})}.
\end{equation}
The scaling factors transform from the unit [\si{\decibel\per\centi\meter\per\mega\hertz}] to a unitless scalar.
The attenuation due to spread is inversely proportional to the distance traveled. We model the scatterers as spheres of radius $r$. The amplitude of the pressure wave then falls off with distance with
\begin{align}
    \attspreadtx & = \frac{r}{d_{TX}}, & \attspreadrx & = \frac{r}{d_{RX}}.
\end{align}
We choose $r=10^{-6}$ to ensure the scatterer radius is significantly smaller than a wavelength.

\subsubsection{Waveform deformation due to frequency dependent attenuation}
\noindent Attenuation as defined in section \ref{sec:methods_attenuation} disregards the fact that attenuation due to absorption is dependent on frequency. High frequency components are attenuated more than low frequency components. This leads to deformations in the waveform. We model this by applying a low-pass filter to the waveform that increases in strength with longer travel times. Let $\opt_a$, and $\opt_b$ be free variables to fit the relation between travel time and low-frequency attenuation. Let $LP(\omega)\{\cdot\}$ denote a low-pass filter with normalized cutoff frequency $\omega$. We then use $LP \left( \opt_a + \opt_b\cdot(\ttx+\trx)\right)\{\wvfm{\tx}{t}\}$ as the transmit waveform. We implement this by pre-computing several low-pass filtered versions of the transmit waveform and applying linear interpolation to find intermediate samples.

\subsubsection{Initial time offset}
\label{sec:methods_initial_time}
\noindent There can be an additional time offset due to the acoustic lens of the system or there can be an offset in the sampled waveform $\wvfm{\tx}{t}$. To compensate for these discrepancies we model a variable time offset in the transmit waveform in $\initialtime$.

\subsubsection{Time gain compensation}
\label{sec:methods_tgc}
\noindent Ultrasound systems apply a \gls{tgc} curve to the incoming signal before analog-to-digital conversion to compensate for the attenuation at greater depths. The \gls{tgc} ensures that the beamformed image does not get darker with depth and reduces quantization noise. However, the \gls{tgc} also distorts the falloff of amplitude. This poses a problem when estimating the attenuation due to element directivity, absorption, and spatial spread (section \ref{sec:methods_directivity}, and \ref{sec:methods_attenuation}). Because of this we need to explicitly include the \gls{tgc} in the model. For this we record the \gls{tgc} curve with the data and include it as a multiplicative factor in the estimate of \gls{rf} data.

\subsection{Optimization}
\label{sec:methods_optimization}
\subsubsection{Stochastic optimization}
Ultrasound \gls{rf} data usually has hundreds of thousands of samples per transmission. This large number of samples makes computing gradients over all samples for every optimization step very computation- and memory-intensive, especially for acquisitions with multiple transmissions. To be able to solve these large problems, we perform \gls{sgd} over batches of samples from the \gls{rf} data generated using (\ref{eq:rf_single}). This approach allows us find solutions even for acquisitions with many transmissions without increasing the computational cost per iteration.

\noindent To obtain our optimization target we compute a mean-squared-error loss between a batch of predicted receive samples $\hat{\mathbf{y}}$ and the corresponding observed samples $\mathbf{y}$.

This loss can be used to apply \gls{sgd} to solve the inverse problem. We optimize this using the ADAM optimizer \cite{kingma_adam_2015}.

\subsubsection{Optimization variables and reparameterization}
We optimize for several model parameters, apart from the scatterers. These are the effective element width $\elw$, the element gain $\elgain$, the initial time offset $\initialtime$, the speed of sound $c$, the absorption coefficient $\attabsorption$, and the parameters for frequency dependent attenuation $\opt_a$, and $\opt_b$.
To constrain the optimization variables to valid ranges we use reparameterization. We denote each optimization variable with a $\opt$. Variables that should be positive are reparameterized with an exponential function. These are the scatterer amplitudes $\allscatamps=\exp \left[ \optvar{\allscatamps}\right]$, and the attenuation coefficient $\attenuation=\exp \left[ \optvar{\attenuation}\right]$.
Variables that should remain within a valid range are reparameterized with a scaled and shifted sigmoid function. These are the speed of sound $c=\sigma \left( \optvar{c}\right)$, and the element gain $\elgain=\sigma(\optvar{\elgain})$. The scaling and shift are tailored to the variable.
Table \ref{tab:reparameterization} shows the reparameterization functions applied to each variable.

\begin{table}
    \centering
    \begin{tabular}{cccc}
        \textbf{name} &\textbf{symbol} & \textbf{reparameterization} &\\\hline\hline
        scatterer amplitude &$\scatamp$ &$\exp(\opt_\scatamp)$\\\hline
        scatterer position &$\scatpos$ &$\frac{1}{\lambda}\cdot \mathbf{\opt}_{\scatpos}$ &\\\hline
        element width&$\elw$ & $\elw_o\cdot\sigma(\opt_\elw)$ \\\hline
        speed of sound  &$c$ &$\exp{(\opt_c)}$ \\\hline
        attenuation & $\attenuation$& $\exp{(\opt_{\attenuation})}$\\\hline
        element gain &$\elgain$& $\frac{1}{2}\cdot(1+\sigma(\opt_\elgain))$ \\\hline
        waveform frequency dependence &$\opt_a$, $\opt_b$& $\exp(\opt_a)$ \\\hline
        $t_0$ offset &$\tz$& $\sigma(\opt_{t_0})$ \\
    \end{tabular}
    \caption{The reparameterization function applied to each optimization variable.}
    \label{tab:reparameterization}
\end{table}

\subsection{Mapping to pixels}
\label{sec:mapping_to_pixels}
\noindent Unlike with regular beamforming algorithms, with \gls{method}, we initially produce a collection of scatterer positions and scatterer amplitudes instead of an image. To visualize this result we must define a mapping to a pixel grid. There are many possible ways to do this. We choose to perform a kernel density estimation by applying a Gaussian window around the scatterers. We define a radius $\plotradius$ of a Gaussian window. The intensity $I$ of a pixel with position $\mathbf{p}$ is then computed as
\begin{equation}
    I(\mathbf{p}, \plotradius) = \sum_{\scatpos_{i}\in \allscatpos} e^{-\left\| \left(\frac{\mathbf{p}-\scatpos_{i}}{\plotradius}\right)^2 \right\|}
\end{equation}
Note that $\plotradius$ is not part of the optimization and can be changed after the solution has been found.\\
Alternative options are binning the scatterer amplitudes to a grid or fitting a 2D function through the scatterers.

\section{Experimental setup}
\label{sec:experimental_setup}

\subsection{Baselines}
\label{sec:baselines}
We compare our method with several pixel-based implementations of classical beamforming algorithms as well as with the \gls{red} regularized beamforming algorithm proposed by Goudarzi et al. \cite{goudarzi_inverse_2022}.

\subsubsection{Signal processing pipeline}
For all beamformers we go through the following steps: First the data is \gls{iq}-demodulated using the Hilbert transform. The data is then shifted to baseband and low-pass filtered to the bandwidth of the transmit waveform using a Butterworth filter. We then perform \gls{tof} correction with lens correction for every pixel in the image individually and apply receive apodization, and an f-number. Specific values are provided in the results section. The resulting complex-valued aperture data vector $\allaperture\in \mathbb{C}^{\NEL}$ is then processed further by the different beamforming algorithms. Finally the beamformed data is compounded, log compressed, normalized by the highest value, and plotted with a dynamic range of $60$\si{\decibel}.

\subsubsection{Delay-and-Sum}
The \gls{das} beamformer produces a beamformed pixel $\beamformedsample$ by simply summing over the \gls{tof} corrected aperture via
\begin{equation}
    \beamformedsample = \sum_{n=0}^{\NEL-1} \aperture_n.
\end{equation}

\subsubsection{Minimum-variance}
\Gls{mv} beamforming is a more advanced and adaptive version of \gls{das}. It achieves significantly better clutter suppression at the cost of higher computational requirements and worse stability because the \gls{mv} beamformer needs to compute the inverse of the autocorrelation matrix of the aperture for every pixel. We implement a more stable version of the \gls{mv} beamformer that applies so-called \textit{spatial smoothing} as presented by Synnevåg et al. \cite{synnevag_adaptive_2007}. This beamformer computes the average correlation matrix over overlapping subapertures of $L$ transducer elements. Let $\mathbf{a}$ be a vector of all ones as the data is \gls{tof}-corrected beforehand, and let $R$ be the averaged correlation matrix over the sub-apertures. The optimal weight $\mathbf{w}$ is then computed as
\begin{equation}
    \mathbf{w} = \frac{R^{-1} \mathbf{a}}{\mathbf{a}^H R^{-1}\mathbf{a}}.
\end{equation}
The beamformed value $\beamformedsample$ is then computed as
\begin{equation}
    \beamformedsample = \frac{1}{\NEL-L+1}\sum_{l=0}^{\NEL-L} \mathbf{w}^H \allaperture_{l:l+L-1}.
\end{equation}
For a more in-depth discussion of the implementation we refer the reader to the original paper by Synnevåg et al. \cite{synnevag_adaptive_2007}.

\subsubsection{Delay-multiply-and-Sum}
The \gls{dmas} beamforming algorithm is a more recent beamformer\cite{matrone_delay_2015} that attempts to reject clutter by pairwise multiplying every pair $\{\aperture_m, \aperture_n|\;m\neq n\}$ before summation. These multiplications amplify coherent signals, while suppressing incoherent signals. To preserve the unit (usually $[\si{\volt}]$), the magnitude of every element in $\allaperture$ is replaced by its square root, while keeping its phase the same. We implement a version of this that is adapted to work in baseband on \gls{iq} data. The beamformed value $\beamformedsample$ is computed via
\begin{equation}
    \beamformedsample = \sum_{n=0}^{\NEL-2}\sum_{m=n+1}^{\NEL-1} \frac{u_n\cdot u_m}{\sqrt{\left\|u_n\right\|}\cdot \sqrt{\left\|u_m\right\|}}.
\end{equation}

\subsubsection{Denoising-based inverse beamforming}
Goudarzi et al. propose to pose imaging as an \textit{inverse beamforming} problem as follows. Let $\mathbf{y}\in \mathbb{R}^{\NAX\cdot\NEL}$ and $\mathbf{x}\in \mathbb{R}^{\NX\cdot\NZ}$ be the flattened channel data and beamformed image respectively. The elements in $\mathbf{x}$ correspond to static grid locations. Let $\Phi$ be a matrix with non-zero elements in the positions where the corresponding pixel has close to the same total \gls{tof} as the sample \cite{goudarzi_inverse_2022}. $\mathbf{v}\in \mathbb{R}^{\NAX\cdot\NEL}$ is a vector of white Gaussian noise samples. The inverse problem is then given by
\begin{equation}
    \mathbf{y} = \Phi \mathbf{x} + \mathbf{w}.
    \label{eq:inverse_beamforming_model}
\end{equation}
For the matrix $\Phi$ we define $\tau_{ax}$ to be the time the sample is recorded, and $\tau_{\pixel}$ is the time the peak of the reflection from a pixel arrives at the element. The matrix $\Phi$ is then defined as
\begin{equation}
    \Phi(\ax, \pixel) = \begin{cases}
        \frac{|\tau_{ax}-\tau_{\pixel}|}{t_{max}} & \text{if }|\tau_{ax}-\tau_{\pixel}| < \frac{1}{f_s} \\
        0                                         & \text{otherwise}
    \end{cases}.
\end{equation}
The model in (\ref{eq:inverse_beamforming_model}) is used to formulate a minimization target. This minimization target is then optimized using \gls{admm}. The authors consider several versions of the method. We will compare to the \gls{red} algorithm, which formulates the following minimization problem:
\begin{equation}
    \argmin_\mathbf{\hat{x}} \left\|\mathbf{y}-\Phi \mathbf{\hat{x}}\right\|^2+\frac{\mu}{2}\cdot \mathbf{\hat{x}}^T \left( \mathbf{\hat{x}}-\mathcal{F}(\mathbf{\hat{x}})\right),
\end{equation}
where $\mathcal{F}(\cdot)$ is a \gls{nlm} denoiser \cite{buades_non-local_2005}. We refer to the original paper for further details on the implementation. We compare against the \gls{red} algorithm with $\mu=2000$ and $\beta=1000$ as suggested by the authors, with a stopping criterion $\epsilon=5\cdot 10^{-4}$ and a smoothing factor $h=0.8$ for the \gls{nlm} denoiser. For acquistions with multiple transmits we apply the method to every transmit individually and then compound the results.
\subsection{Evaluation metrics}
\label{sec:methods_metrics}
\subsubsection{Generalized contrast to noise ratio}
The \gls{gcnr} measures how much overlap exists between the histograms of pixel intensities between two regions \cite{rodriguez-molares_generalized_2020} . If two regions share no pixel intensities, the \gls{gcnr} is $1.0$. If both regions have the exact same intensity distribution the \gls{gcnr} is $0.0$. We evaluate the \gls{gcnr} with $256$ bins.

\subsubsection{Mean squared error}
We evaluate the error in \gls{rf} data based on the \gls{mse} between the recorded \gls{rf} data and the reconstructed \gls{rf} data. Since this data is amplified by the \gls{tgc} gain, we can expect deeper regions to have a contribution that is similar to the contribution of the closer regions.

\subsection{Evaluation data}
\label{sec:data}
\noindent We evaluate our method on data with a Verasonics Vantage 256 research ultrasound system. The data was captured with a 128-element linear transducer (Verasonics L11-5V), and with an 80-element phased-array transducer (Philips S5-1). The \gls{rf} data are sampled at double the Nyquist frequency and low-pass filtered before analog-to-digital conversion.
We acquired data with a tissue mimicking phantom (CIRS040) to have a simple target with known properties. Additionally we acquired \textit{in-vivo} data. These are scans of a carotid artery with the linear transducer and cardiac data with the phased array transducer.

\section{Results}
\label{sec:results}
\subsection{Tissue mimicking phantom with phased array probe}
\label{sec:results-cirs-phantom}
\subsubsection{Diverging-wave acquisition}
\begin{figure}
    \centering
    \includegraphics[width=\linewidth]{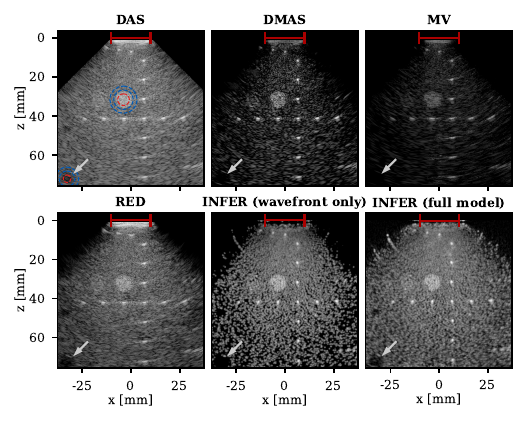}
    \caption{Single diverging wave acquisition of a CIRS phantom beamformed with \gls{das}, \gls{mv}, \gls{dmas}, \gls{red}, and \gls{method}. "wavefront only" and "full model" are the reconstructions for \gls{method} based on (\ref{eq:model_wavefrontonly}), and (\ref{eq:model_allelements}) respectively ([-70, 0]dB).}
    \label{fig:cirs_plane_wave}
\end{figure}

\noindent Fig. \ref{fig:cirs_plane_wave} shows the reconstruction of \gls{method} for a single-diverging-wave acquisition based on the \textit{full model} in (\ref{eq:model_allelements}) and the \textit{wavefront only model} in (\ref{eq:model_wavefrontonly}). The baselines were all run with an f-number of $0.5$. For the \gls{mv} beamformer we use a subaperture size of $30$ and apply a diagonal loading of $1\cdot 10^{-4}$. For \gls{red} we use a grid of $1226\times307$ grid points. For \gls{method} we use an initial grid of $384\times 256$ scatterers. The full model predicted a speed of sound of $\SI{1533}{\meter/\second}$. The wavefront only model predicted a speed of sound of $\SI{1539}{\meter/\second}$. The stated speed of sound in the phantom is $\SI{1540}{\meter/\second}$. The resulting \gls{gcnr} for this and the following experiments are shown in table \ref{tab:gcnr}.
The results produced by \gls{method} show an increase in resolution, especially at greater depth. The hypoechoic cyst in the lower left corner of the image shows reduced clutter in the both \gls{method} results. Compared to \gls{mv}, and \gls{dmas} we see a better preservation of the background echogenicity.

\subsubsection{Synthetic aperture transmit scheme}
\noindent For this experiment we choose to use a somewhat unusual scheme where we do a sparse synthetic aperture acquisition. We acquire data from 2 transmissions where only a single element fires and all elements receive. With this acquisition scheme the \textit{full model} is (\ref{eq:model_allelements}) identical to the \textit{wavefront only} model. This means we can have a small model error without a large computational burden.

\begin{figure}
    \centering
    \includegraphics[width=\linewidth]{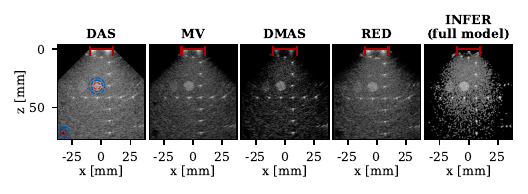}
    \caption{2 transmit acquisition of a CIRS phantom reconstructed with \gls{das}, \gls{mv}, \gls{dmas}, and \gls{method}.}
    \label{fig:cirs_synthetic_aperture}
\end{figure}

Fig. \ref{fig:cirs_synthetic_aperture} shows the images reconstructed using \gls{das}, \gls{mv}, \gls{dmas}, \gls{red}, and \gls{method} respectively. All results were produced from the same $2$ single-element acquisitions. We observe better resolution in the \gls{method} result both in the middle and on the sides of the image. The \gls{method} result fails to capture the background echogenicity at greater depth. However, it does resolve the wire targets to the side around $x=25\si{\centi\meter}$ more clearly than all other methods.

\subsection{In-vivo data with linear probe}
\subsubsection{Synthetic aperture acquisition}
\begin{figure*}
    \centering
    \includegraphics{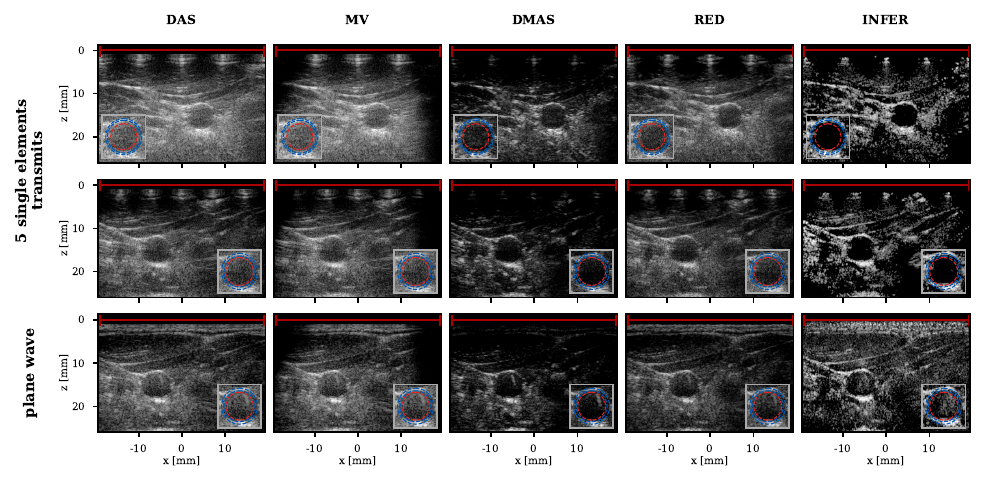}
    \caption{Reconstruction of two cross-sectional views of a carotid artery with \gls{das}, \gls{mv}, \gls{dmas}, \gls{red} and \gls{method} ([-60, 0]dB). The used transmit scheme is 5 single-element transmissions in the top and middle rows and plane wave in the bottom. The red line indicates the transducer aperture.}
    \label{fig:carotid_sa2}
\end{figure*}

\noindent Here we evaluate the performance of the methods on a cross-sectional acquisition of the carotid artery with 5 \gls{sa} transmissions. We run all baselines with an f-number of $2.5$. For the \gls{mv} beamformer we use a subaperture size of $30$ and apply a diagonal loading of $1\cdot 10^{-4}$. For \gls{method} we show results for the \textit{full model} with an initial grid of $320\times 256$ scatterers. The speed of sound estimated by the model was \SI{1553}{\meter/\second}.\\
Fig. \ref{fig:carotid_sa2} shows reconstructions produced by all methods. We compute the \gls{gcnr} between the interior of the artery and the arterial wall. The top reconstruction shown in the top row predicted a speed of sound of $\SI{1548}{\meter/\second}$. The reconstruction shown in the middle row predicted a speed of sound of $\SI{1555}{\meter/\second}$. We observe a significant increase in \gls{gcnr} with \gls{method}. The images produced by \gls{das}, \gls{mv}, and \gls{red} all contain strong clutter artifacts, which are especially apparent in the near field and on the sides of the images. \gls{dmas} suffers less from these artifacts, but suppresses a lot of structure in the image. \gls{method} does not suffer from these clutter artifacts while reconstructing the structure well across the image.

\subsubsection{Plane wave acquisition with full model}
The bottom row of Fig. \ref{fig:carotid_sa2} shows the results on a 3 plane wave acquisition of a carotid artery. The \gls{method} results are generated using the \textit{full model}. The results show more clutter across the image, especially in the artery. The obtained \gls{gcnr} as shown in table \ref{tab:gcnr} is significantly lower than in the results in Fig. \ref{fig:carotid_sa2}.

One possible explanation for this worse performance is that the speed of sound likely varies significantly across the image. The muscle close to the transducer has a high speed of sound, while the speed of sound in the vessel will be smaller. This in combination with the complexity introduced by the $128$ wavefronts could make the model more inaccurate than in the phantom results in Fig. \ref{fig:cirs_plane_wave} or the synthetic aperture results in Fig. \ref{fig:carotid_sa2}.

\subsection{Cardiac acquisition}
\begin{figure}
    \centering
    \includegraphics[width=\linewidth]{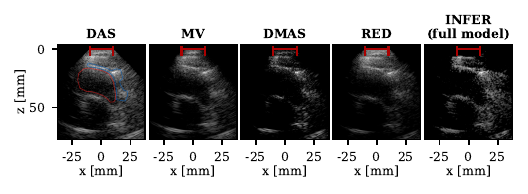}
    \caption{Reconstruction of a diverging wave transmit acquisition of an apical 2-chamber view of a heart ([-60, 0]dB). The reconstruction was performed based on three angled diverging wave transmits.}
    \label{fig:cardiac}
\end{figure}

\noindent Fig. \ref{fig:cardiac} shows the reconstruction of a diverging wave transmit cardiac acquisition of an apical 2-chamber view. The \gls{method} reconstruction was performed with an initial grid of $256\times 256$ scatterers. The reconstruction was performed based on three diverging wave transmits with angles $[-3.6^\circ, 0.0^\circ, 5.4^\circ]$. The results show that \gls{method} reduces haze in the near field without suppressing tissue as much as \gls{dmas}, and \gls{red}. Table \ref{tab:gcnr} shows the \gls{gcnr} between the ventricle and the cardiac wall. \gls{method} effectively suppresses the haze in the top of the image. An explanation might be that multipath signal components do not fit any tissue location in the model, causing them to be rejected.

\begin{table}[]
\resizebox{\linewidth}{!}{\begin{tabular}{lllllll}
& \textbf{DAS}  & \textbf{MV}   & \textbf{DMAS} & \textbf{RED}  & \makecell{\textbf{\gls{method}} \\ \textbf{(full)}}& \makecell{\textbf{\gls{method}}\\\textbf{(wfnt only)}} \\ \hline \hline
\multicolumn{1}{l}{\makecell{phantom DW\\cyst hyper}}         & \textbf{0.82} & 0.80 & 0.71 & 0.61 & 0.80& \textbf{0.82}\\\hline
\multicolumn{1}{l}{\makecell{phantom DW\\cyst hypo}}         & 0.66 & 0.71 & 0.58 & 0.55 & \textbf{0.85}& 0.72\\\hline

\multicolumn{1}{l}{\makecell{phantom SA\\cyst hyper}}         & 0.79 & 0.82 & 0.80 & 0.61 & \textbf{0.94}&-\\\hline
\multicolumn{1}{l}{\makecell{phantom SA\\cyst hypo}}         & 0.61 & \textbf{0.64} & 0.60 & 0.53 & 0.45& -\\\hline

\multicolumn{1}{l}{carotid SA 1} & 0.58 & 0.58 & 0.64 & 0.58 & \textbf{0.93} & - \\\hline
\multicolumn{1}{l}{carotid SA 2} & 0.58 & 0.59 & 0.64 & 0.64 & \textbf{0.79} & - \\\hline
\multicolumn{1}{l}{carotid pw}         & 0.49 & 0.49 & 0.49 & 0.45 & \textbf{0.52} & - \\\hline
\multicolumn{1}{l}{cardiac dw}         & 0.38 & 0.37 & 0.41 & 0.28 & \textbf{0.45} & - \\

\end{tabular}}
\caption{\gls{gcnr} results for all shows acquisitions. The best values are in bold.}
\label{tab:gcnr}
\end{table}

\subsection{RF reconstruction error}
\begin{figure}
    \centering
    \includegraphics{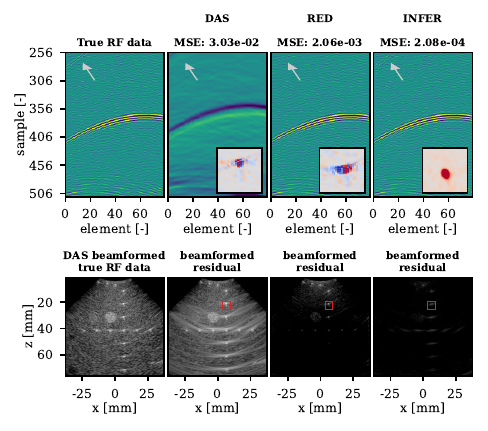}
    \caption{Comparison of the RF data reconstructed from the images of different methods. The top row shows a section of the \gls{rf} data with the corresponding image before log compression, revealing oscillations in the \gls{das}, and \gls{red} solutions. The bottom row shows the images obtained from \gls{das} beamforming the reconstructed \gls{rf} data in the top row.}
    \label{fig:rf_data_comparison}
\end{figure}

\noindent In this section we investigate the extent to which the method is able to faithfully reconstruct the measured \gls{rf} data. For \gls{method} and \gls{red} we can just project the solutions back into the \gls{rf} domain using their respective measurement models. We also show how the \gls{das} solution maps to the \gls{rf} domain using the \gls{method} measurement model. The top row of Fig. \ref{fig:rf_data_comparison} shows a part of the \gls{rf} data for each of these along with the \gls{mse} fitting error. The inset plot shows the reflectivity estimate produced by all methods before log compression. The regions we focus on here correspond to a single wire target, indicated by the red box.

Interestingly the \gls{red} solution does manage to fit the full waveform shape in the \gls{rf} data well, even though it models the waveform shape as a triangle pulse that is fully non-negative. The \gls{red} solution is able to do this by placing both positive and negative scatterers around a source of backscatter. We see this in the inset plot of the pre-log compressed image. By doing so, the \gls{red} algorithm builds up the waveform and fits the \gls{rf} data, but does so with a solution that has no clear physical interpretation. \gls{method} produces a closer fit with only positive backscatter amplitudes.

The bottom row of Fig. \ref{fig:rf_data_comparison} shows the images obtained from beamforming these residuals. These beamformed residual images can show which regions in the image have been properly fitted by the model and which regions still contain structure. As expected the results show that the \gls{das} beamformer does not produce an image that reconstructs the \gls{rf} data well. We see that \gls{method} achieves a lower \gls{mse} and leaves less structure in the residual image than both \gls{das}, and \gls{red}. However, \gls{red} seems to be able to fit the far field scatterers, while \gls{method} does leave some structure there.

\subsection{Ablation study}

\begin{table}
    \centering
    \begin{tabular}{ccc}
        \textbf{Ablated Feature} & \textbf{RF MSE} & \textbf{delta MSE} \\\hline\hline
        \\[-1em]
        None &  $2.64 \times 10^{-4}$ & $0.0 \times 10^{-4}$\\\hline
        \\[-1em]
        Element Directivity &  $2.71 \times 10^{-4}$ & $+0.07 \times 10^{-4}$\\\hline
        \\[-1em]
        Element Gain &  $2.82 \times 10^{-4}$ & $+0.18 \times 10^{-4}$\\\hline
        \\[-1em]
        Attenuation from Spread &  $2.48 \times 10^{-4}$ & $-0.16 \times 10^{-4}$\\\hline
        \\[-1em]
        Attenuation from Absorption &  $2.66 \times 10^{-4}$ & $+0.02 \times 10^{-4}$\\\hline
        \\[-1em]
        Waveform Deformation &  $2.64 \times 10^{-4}$ & $0.0 \times 10^{-4}$\\\hline
        \\[-1em]
        Initial Time Offset &  $2.64 \times 10^{-4}$ & $0.0 \times 10^{-4}$\\\hline
        \\[-1em]
        Time Gain Compensation &  $4.47 \times 10^{-4}$ & $+1.83 \times 10^{-4}$\\
        
    \end{tabular}
    \caption{The Mean Squared Error (MSE) achieved with ablated features with the delta to None.}
    \label{tab:ablation-results}
\end{table}

\begin{figure}
    \centering
    \includegraphics[width=\linewidth]{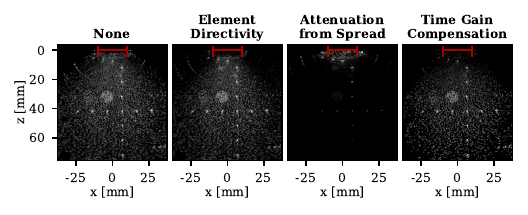}
    \caption{Images produced by models with ablated features. The title of each plot indicates which feature was ablated, with \textit{None} being the full unablated model.}
    \label{fig:ablation-partial}
\end{figure}

\begin{figure}
    \centering
    \includegraphics[width=0.9\linewidth]{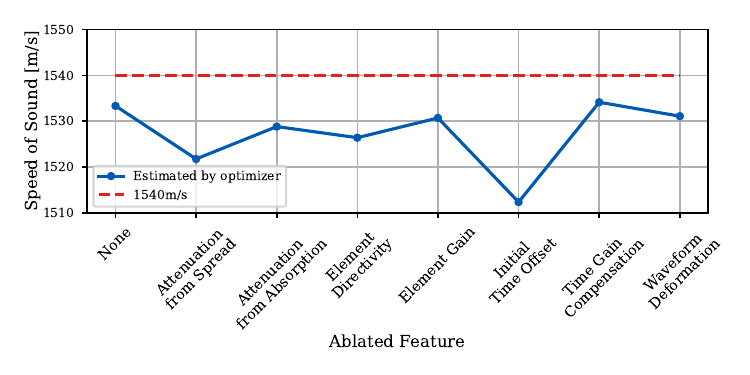}
    \caption{The estimated speed of sound under each ablation.}
    \label{fig:speed-of-sound}
\end{figure}

\noindent \gls{method} aims to produce accurate images of the anatomy by correctly modelling a number of physical phenomena relating the anatomy to the observed RF data, as described in Section \ref{sec:methods}. Ideally, these modelling features should be physically accurate, leading to higher image quality and correctness. In order to evaluate the relevance of each feature, we carried out an \textit{ablation study}, wherein we removed or `ablated' each feature one-by-one, and solved the inverse problem with the resulting set of partial models. The aim of such a study is to measure the importance of each parameter through the impact that removing it has on the results. We fit each ablated model to the same \gls{rf} data, from the acquisition on the CIRS phantom described in Section \ref{sec:results-cirs-phantom}. This resulted in fitting 8 models in total, one for each of the features described in Sections \ref{sec:methods_directivity} through \ref{sec:methods_tgc}, and one model with no ablated features, denoted by \textit{None} in the figures. It is clear in Fig. \ref{fig:ablation-partial} that modelling attenuation is essential to ensuring the visibility of deeper regions of tissue. The impact of modelling element directivity can also be seen in Fig. \ref{fig:ablation-partial}: without it, peripheral regions become muted. The RF data reconstruction error for each model is provided in Table \ref{tab:ablation-results}. Most notable is the error caused by removing \gls{tgc}, indicating that \gls{tgc} is essential for accurate RF reconstruction. The rest of the results are quite similar, typically matching or slightly degrading the reconstruction relative to the full model. One exception is in removing Attenuation from Spread, which slightly improved the \gls{rf} reconstruction error, but at the expense of image quality, as seen in Fig. \ref{fig:ablation-partial}. We observe in Fig. \ref{fig:speed-of-sound} that physical parameters, such as speed of sound, can vary significantly according to which features are ablated. The CIRS phantom should have a speed of sound of $1540$ \si{\meter/\second}, which is estimated reasonably well by the full model, but quite poorly when certain features are removed, for example, Initial Time Offset. Indeed, if certain physical phenomena are not explicitly represented in the model, the optimizer will find other ways to account for those phenomena using the parameters available to it, in order to minimize the loss. While this may result in similar or even reduced \gls{rf} reconstruction error, the correctness of the physical parameter estimates can become compromised. Future work might look therefore towards evaluating the correctness of these physical parameters in cases where the true values are known, supporting the case for including such information estimated by inverse scattering models such as \gls{method} in diagnostic tasks.

\section{Discussion}
\label{sec:discussion}
\noindent The results show that \gls{method} is able to reconstruct images across different imaging targets and transmit schemes. The method works both on a tissue mimicking phantom and on \textit{in-vivo} data. \gls{method} especially compares favourably in conditions where off-axis clutter or multi-path signals dominate, like in the carotid artery or in the near field in cardiac acquisitions. We suspect this is because the model can either attribute these signal components to their actual source, or it cannot attribute these to any location and thus rejects them.

All the results we have shown have been produced without adding a regularization term in the optimization target. The regularization due to the implicit inductive bias of the \gls{sgd} optimization and constraints like the limited number of scatterers already seem to provide a well-posed problem. Still, more sophisticated modelling of priors over the space of solutions might further improve the image quality and convergence speed.

It is important to note that the method presented in this paper does not preserve the texture of spatially correlated speckle as this is a result of the beamforming process. Regularized beamforming algorithms like \gls{red} employ the classical beamforming behaviour of summing over hyperbolas of equidistant points. As such, these regularized beamforming algorithms reproduce correlated speckle in the images. Inverse scattering approaches like \gls{method} do not do so. This means they should produce images with only spatially uncorrelated multiplicative noise, given that the forward model is fully correct and the optimization procedure successful. This lack of correlated speckle means that traditional speckle tracking methods might not work correctly. However, while some established methods may not work with inverse-scattering-based imaging, some new possibilities become available. Motion tracking could potentially be performed by initializing the optimization for frame $k$ with the result of frame $k-1$. We could then track the movement of every scatterer between frames.

Another point that should be noted is the runtime for the presented method. The presented method is currently too slow to be applied in real-time. Solving for an image currently takes anywhere between several minutes to several hours, depending on the acquisition scheme, hardware and on parameters like the batch size and whether the \textit{full model} is used or not. There are several ways the runtime could be improved. We could likely achieve significant gains in speed by initializing every next frame with the solution from the previous frame. Another possibility is to divide the image into smaller regions and solve these in parallel. It has also been shown that machine learning based methods can enable acceleration of model-based methods.

We have shown that flaws in modelling the acquisition process lead to reduced image quality. This suggests that improving the physical forward model could improve image quality even further. Potential targets include modeling quantities like the speed of sound and absorption coefficient locally instead of globally, better modelling of the acoustic lens and the wave dynamics in the near field, hard reflections, and multiple scattering. The challenge with some of these is that they introduce dependencies that violate the assumption that the \gls{rf} data is just a linear superposition of independent scatterer responses.

\section{Conclusions}
\label{sec:conclusion}
\noindent We have presented \gls{method}: an off-grid stochastic optimization algorithm for the inverse scattering problem in ultrasound imaging. We have shown that \gls{method} is able to produce high-resolution images with strongly reduced clutter while accurately fitting the \gls{rf} data. The results show that \gls{method} compares favorably against other (regularized) beamforming methods, across various imaging targets (in-vitro and in-vivo) and acquisition schemes.
With \gls{method}, we have shown that it is possible to find plausible solutions to the inverse scattering problem in ultrasound imaging without employing priors for regularization that introduce bias.
This method may pave the way towards standardized and quantitative ultrasound imaging that is grounded in the physics of the acquisition process.

\bibliographystyle{IEEEtran} 
\bibliography{references} 

\end{document}